\documentclass[a4paper,11pt]{article}
\usepackage{jinstpub} 
\usepackage{lineno}
\usepackage{siunitx}
\usepackage{graphicx}
\usepackage{caption}
\usepackage{subcaption}
\usepackage{xcolor}
\DeclareUnicodeCharacter{0301}{\hspace{-1ex}\'{ }}

\title{\boldmath Gain layer degradation study after neutron and proton irradiations in Low Gain Avalanche Diodes}

\author[a,1]{E. Curr\'as Rivera,\note{Corresponding author.}}
\author[a]{A. La Rosa,}
\author[a]{M. Moll}
\author[b]{and F. Zareef}
\affiliation[a]{CERN, Organisation europ\'enne pour la recherche nucl\'eaire,\\
CH-1211 Genéve 23, Switzerland}
\affiliation[b]{Faculty of Physics and Applied Computer Science \\
AGH-University of Science and Technology \\
al. Mickiewicza 30 30-059 Krakow, Poland}
\emailAdd{esteban.curras.rivera@cern.ch}

\abstract{ The high-luminosity upgrade of the ATLAS and CMS experiments  includes dedicated sub-detectors to perform the time-stamping of minimum ionizing particles (MIPs). These detectors will be exposed up to fluences in the range of $1.5-2.5 \times 10^{15}\,n_{eq}/cm^2$ at the end of their lifetime and, Low Gain Avalanche Diode (LGAD) has been chosen as their baseline detection technology. To better understand the performance of LGAD detectors in these environments, a gain layer degradation study after neutron and proton irradiations up to a fluence of $1.5\times10^{15}\,n_{eq}/cm^2$ was performed. LGADs manufactured at Hamamatsu Photonics (HPK) and Centro Nacional de Microelectr\'{o}nica (CNM-IMB) were chosen for this study and, a comparison in the gain layer degradation after exposure to reactor neutrons at the Jožef Stefan Institute (JSI) in Ljubjana and \mbox{24 GeV/c} protons at the CERN-PS is presented here.}


\begin{document}
\maketitle
\flushbottom

\section{Introduction}
\label{sec:intro}

The high-luminosity upgrade of the Large Hadron Collider (HL-LHC) is foreseen to start operation at the beginning of 2029 delivering an integrated luminosity of up to $4000\,fb^{-1}$ during its 10\,years of operation. The HL-LHC will operate at a stable luminosity of $5.0\times10^{34}$\,$cm^{-2}s^{-1}$, with an ultimate scenario of $7.5\times10^{34}$\,$cm^{-2}s^{-1}$ \cite{Aberle:2749422}. For this upgrade, the ATLAS and CMS experiments will include dedicated sub-detectors to perform timing measurements of minimum ionizing particles (MIPs) during the HL-LHC operations \cite{CERN-LHCC-2017-027,CERN-LHCC-2018-023}.\

\begin{figure}[h]
    \centering
    \includegraphics[width=0.46\textwidth]{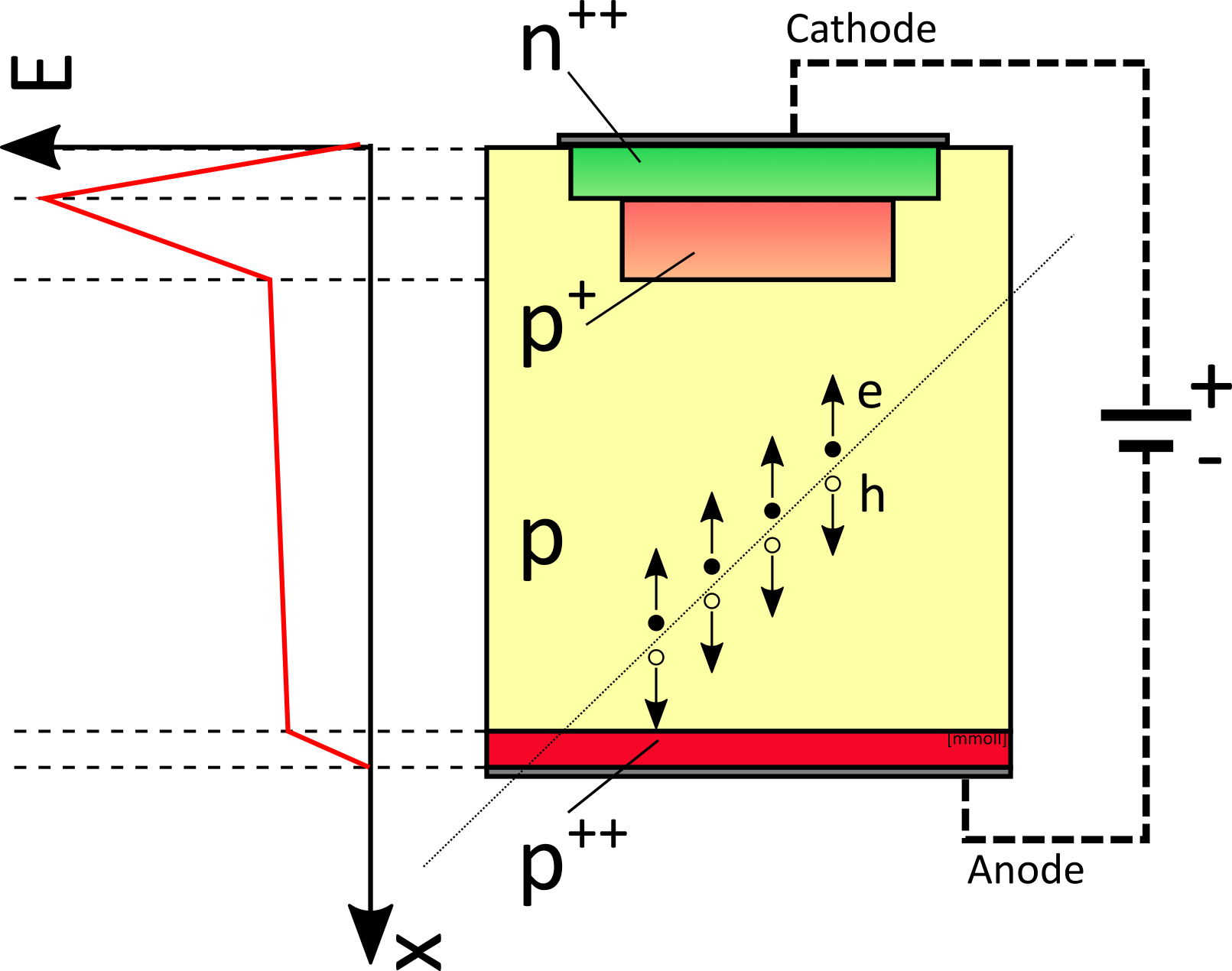}
    \includegraphics[width=0.52\textwidth]{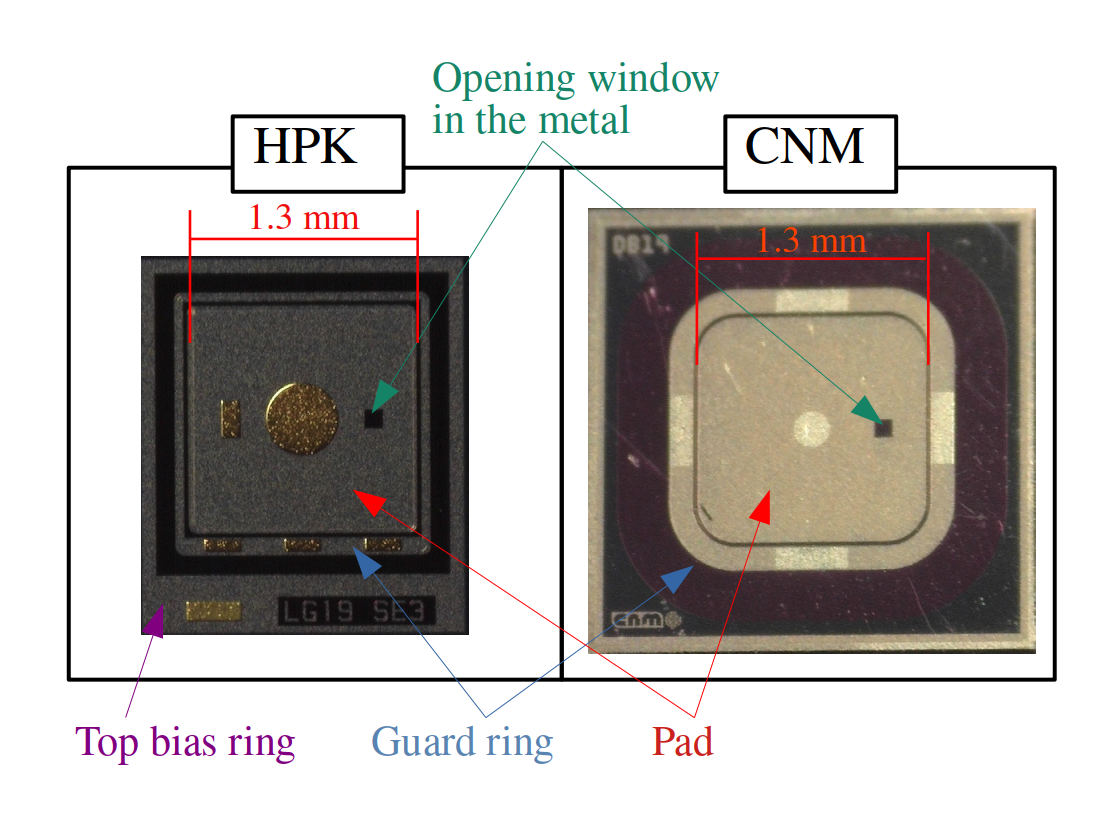}
    \caption{Schematic diagram of a standard LGAD. Photographs (not scaled) of the two LGAD types studied in this work.}
    \label{figure:1}
\end{figure}

The MIP timing sub-detector sensors will be made of Low Gain Avalanche Diodes (LGAD) \cite{PELLEGRINI201412,CARTIGLIA2015141,m2019silicon}. LGADs are semiconductor detectors with signal amplification, that are implemented as $n^{++}-p^+-p$ avalanche diodes. The highly-doped $p^{+}$ layer is added to create a very high electric field region. This electric field generates the avalanche multiplication of the primary electrons, creating additional electron-hole pairs. The schematic cross-section of a standard pad-like LGAD is shown in figure\,\ref{figure:1} (left). The LGAD structure is designed to exhibit a moderate gain and operate over a wide range of reverse bias voltages before breakdown.

ATLAS and CMS MIP timing sub-detectors are proposed to be built using LGADs with a pixel size of $1.3\times1.3\,mm^2$. These detectors will be exposed to radiation levels up to $2.5\times10^{15}\,n_{eq}/cm^2$ (ATLAS HGTD) and $1.5\times10^{15}\,n_{eq}/cm^2$ (CMS Endcap), posing a major challenge for this technology. The radiation fields are mainly composed of neutrons but also contain charged particles in different ratios along the detector dimensions, thus studying the radiation damage produced by all types of particles involved is crucial to understand the performance of LGADs during the HL-LHC operation. Although similar studies have been conducted with other types of LGADs, a direct comparison between neutrons and \mbox{24 GeV/c} protons has not been performed yet with the LGADs studied in this work. Most studies in the literature focus primarily on neutrons \cite{CERN-LHCC-2020-007, CMS:2667167}.

\section{Samples description}

The samples used for this study were LGAD and PIN sensors produced by Hamamatsu Photonics (HPK) and Centro Nacional de Microelectr\'{o}nica (CNM-IMB). The LGAD and PIN sensors from the respective producers differed only in the addition of the $p^+$-implant, i.e. the gain layer (GL), for LGADs.  The HPK samples were from the production run S10938-6130 (also called HPK prototype 2 or HPK2) produced on a wafer with a \SI{50}{\micro\meter} epitaxial layer on a \SI{150}{\micro\meter} thick low resistivity support wafer. The CNM samples were from the production run 12916, produced on a \SI{50}{\micro\meter} Float Zone wafer bonded to a \SI{300}{\micro\meter} low resistivity Czochralski wafer as support. The CNM LGADs were designed with a shallow gain layer doping profile while the HPK ones have a deep gain layer doping profile. 

All samples have an active area of \mbox{1.3$\times$1.3\,mm$^2$} and a guard ring structure surrounding the central pad. To allow laser illumination from the pad side (i.e. the front electrode), they have an opening window  of \mbox{100$\times$\SI{100}{\micro\meter}$^2$} in the metallization. In \mbox{Fig.\,\ref{figure:1}}, two pictures of sensors studied in this work are shown, and in Table\,\ref{table_1} the key parameters of the samples are listed.

\begin{table}[htbp]
\begin{center}
\caption{Main parameters for the LGAD samples used in this work: full depletion voltage (V$_{dep}$), gain layer depletion voltage (V$_{gl}$), average breakdown voltage (V$_{bd}$) at 20$^\circ$C, capacitance reached above full depletion (C$_{end}$) and active thickness ($d$).\\}
\label{table_1}
\begin{tabular}{| l | c | c | c | c | c |}
\hline
\textbf{Sample} & \textbf{V$_{dep}$ [V]} & \textbf{V$_{gl}$ [V]} & \textbf{V$_{bd}$\,(20$^\circ$C) [V]} & \textbf{C$_{end}$}  [\si{\pico\farad}]& \textbf{\textit{d} [\si{\micro\meter}]} \\
\hline
HPK2-W25 & 61.7  & 54.5  & 145  & 3.6  & 48 \\
CNM    & 42.8  & 39.4  & 80 - 100  & 4.2  & 42 \\
\hline
\end{tabular}
\end{center}
\end{table}

\begin{figure}[h]
    \centering
    \includegraphics[width=0.46\textwidth]{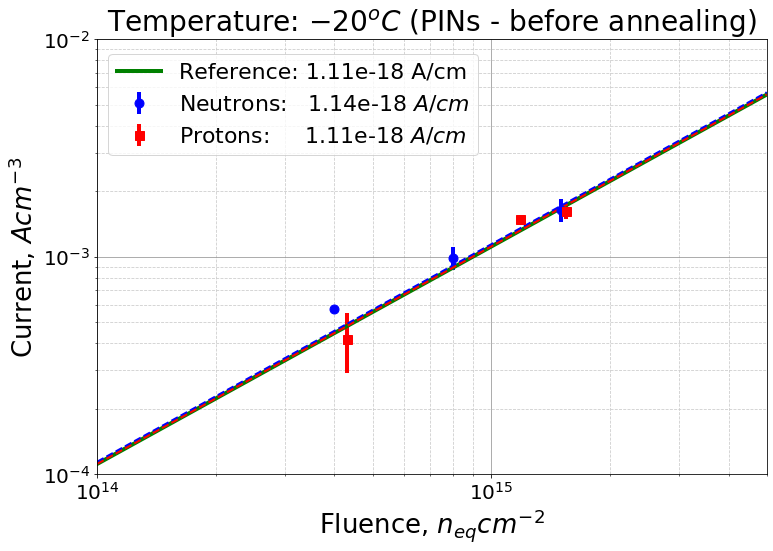}
    \caption{Pad current as a function of the fluence, measured with the PIN diodes at $-20\,^oC$ before annealing. Dotted lines represent the linear fit to the data. The reference value is included for comparison \cite{Moll:1999kv}.}
    \label{figure:2}
\end{figure}

A subset of the samples was irradiated at the CERN-PS irradiation facility with \mbox{24\,GeV/c} protons \cite{Gkotse:2015axt} at 3 different fluences. The applied hardness factor for conversion into 1 MeV neutron equivalent damage was $~0.62$ \cite{Allport_2019}, and the fluences achieved were measured within an error of $7\,\%$. Another subset of samples was irradiated with neutrons in the TRIGA II reactor at the Jožef Stefan Institute (JSI) in Ljubljana \cite{1999NIMPA.426...51Z} at 3 different fluences, achieved within an error of $10\,\%$. Some PIN diodes and LGADs were kept unirradiated for reference.
In table\,\ref{table_1} a summary of all LGADs available and all the irradiation fluences is shown. Due to a technical problem during the proton irradiation, the medium fluence was higher than the requested one of  $8.0\times10^{14}$\,$n_{eq}cm^{-2}$. The normalized bulk current as a function of the fluence, extracted from the irradiated PIN detectors after full depletion, is shown in \mbox{figure\,\ref{figure:2}}. Both irradiations, protons, and neutrons, are in good agreement with the expected value \cite{Moll:1999kv}.

\begin{table}[t]
\begin{center}
\caption{Table summarizing the irradiations performed in the different LGADs and PINs studied in this work.\\} 
\label{table_2}
 \begin{tabular}{| l | c | c | c |} 
 \hline
 Particle & Low fluence & Medium fluence & High fluence \\ [0.7ex] 
  & [$n_{eq}/cm^2$] & [$n_{eq}/cm^2$] &  [$n_{eq}/cm^2$] \\
 \hline\hline
 Proton & $4.3\times10^{14}$ & $1.18\times10^{15}$  & $1.55\times10^{15}$ \\[0.5ex] 
 \hline
  Neutron & $4.0\times10^{14}$ & $8.0\times10^{14}$  & $1.5\times10^{15}$ \\[0.5ex] 
 \hline
\end{tabular}
\end{center}
\end{table}

\section{Electrical characterization}

\begin{figure}[!t]
    \centering
      \begin{subfigure}[b]{0.49\textwidth}
         \centering
         \includegraphics[width=\textwidth]{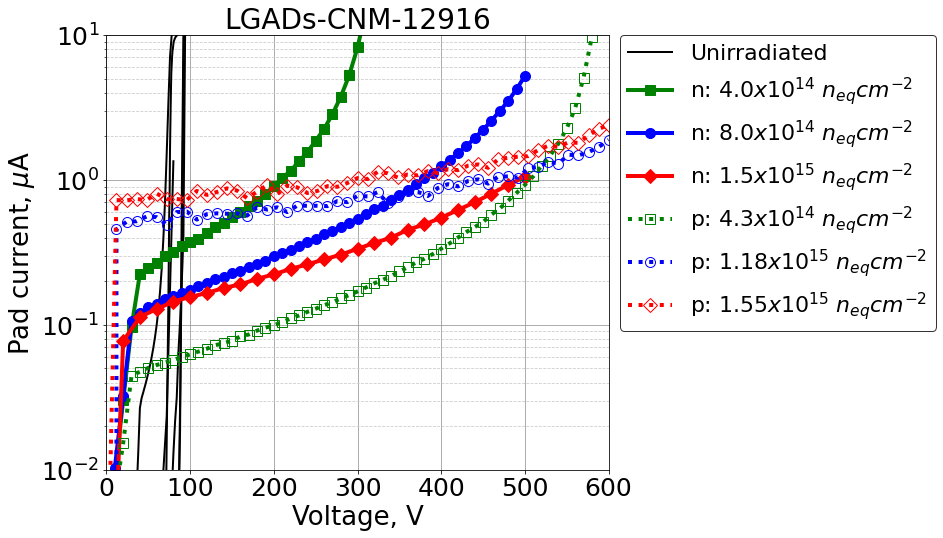}
         \caption{}
         \label{}
     \end{subfigure}
     \hfill
     \begin{subfigure}[b]{0.49\textwidth}
         \centering
         \includegraphics[width=\textwidth]{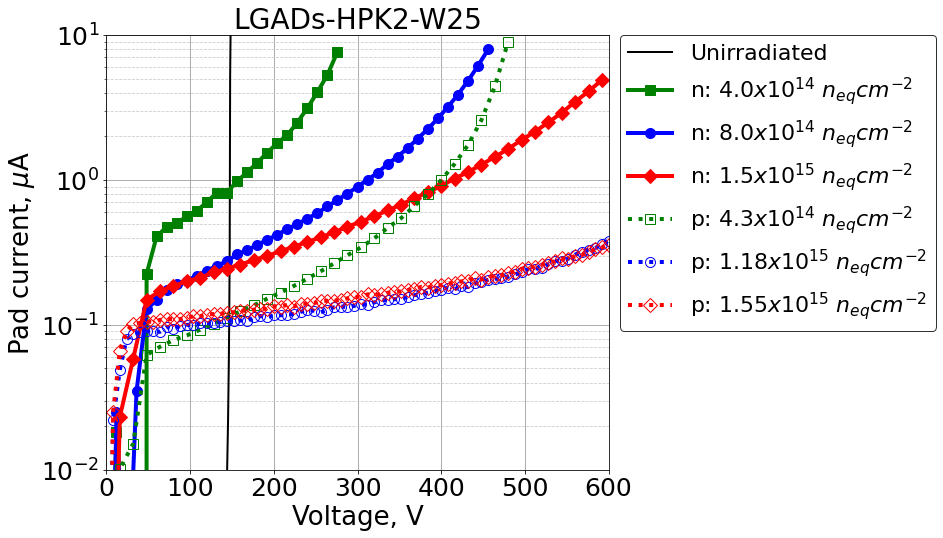}
         \caption{}
         \label{}
     \end{subfigure}
     \hfill
    \caption{Pad current of the LGADS studied in this work. The figure shows the measurements of the CNM LGADs (a) and the HPK ones (b).}
    \label{figure:3}
\end{figure}

The electrical characterization was performed at the SSD labs at CERN. A probe station to measure the bare samples before and after irradiation was used. The samples were placed directly on a chuck that can be temperature controlled. The leakage current and the capacitance as a function of the reverse bias voltage were measured after an annealing step of 80 minutes at $60\,^oC$. During the electrical characterization, the guard ring was always grounded.
Before irradiation, the leakage current was measured at $20\,^oC$, while after irradiation the temperature was set to $-20\,^oC$. In \mbox{figure\,\ref{figure:3} (a)} the pad current as a function of the reverse bias for the CNM LGADs is shown, and in \mbox{figure\,\ref{figure:3} (b)} for the HPK LGADs.

\begin{figure}[!t]
    \centering
  \begin{subfigure}[b]{0.49\textwidth}
         \centering
         \includegraphics[width=\textwidth]{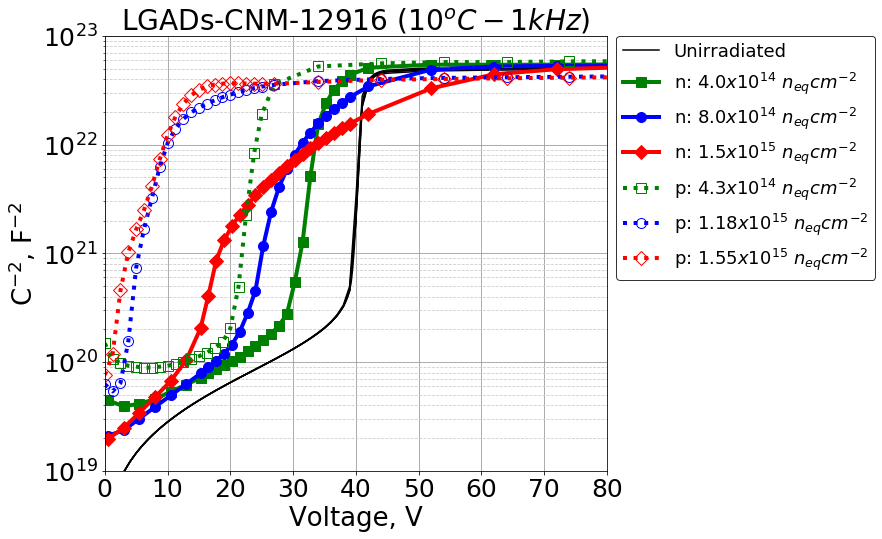}
         \caption{}
         \label{}
     \end{subfigure}
     \hfill
     \begin{subfigure}[b]{0.49\textwidth}
         \centering
         \includegraphics[width=\textwidth]{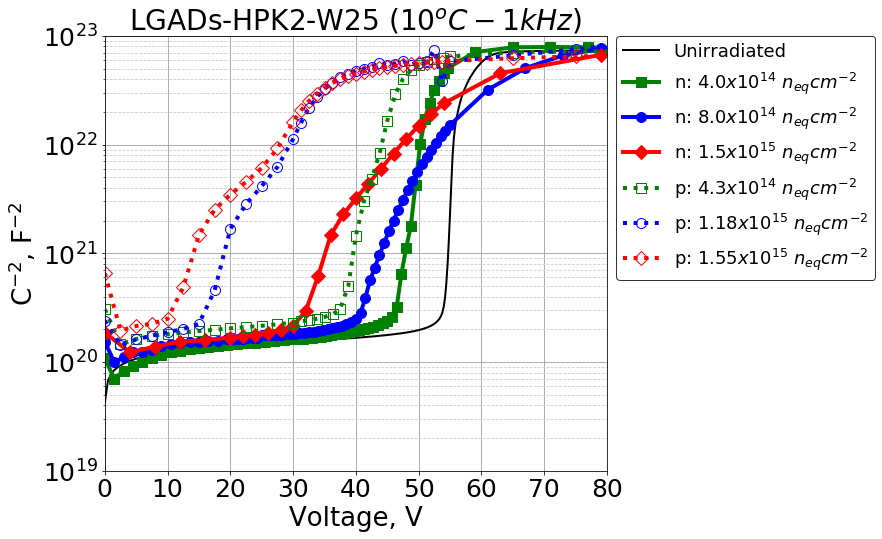}
         \caption{}
         \label{}
     \end{subfigure}
     \hfill
    \caption{Capacitance characterization of the LGADs studied in this work. The figure shows the inverse of the squared capacitance as a function of the bias voltage for the CNM LGADs (a) and the HPK ones (b).}
    \label{figure:4}
\end{figure}

The capacitance before irradiation was measured at $20\,^oC$, and the LCR meter frequency was set to 1\,kHz. After irradiation, the temperature was set to $10\,^oC$, and the LCR meter frequency was kept at 1\,kHz. In \mbox{figure\,\ref{figure:4} (a)} the capacitance as a function of the reverse bias for the CNM LGADs is shown, and in \mbox{figure\,\ref{figure:4} (b)} for the HPK LGADs.

During the electrical characterization, a compliance of $10\,\mu A$ in the total current  was set and the bias voltage was kept always below \mbox{600 V}.

\section{Infrared laser measurements}

The gain of the devices was measured with the Transient Current Technique (TCT) setup of the SSD lab at CERN, using a pulsed IR-laser of \mbox{1060 nm} and a pulse width of \mbox{250 ps}. The LGADs are glued on a customized Printed Circuit Board (PCB) that is placed on top of a temperature-controlled metallic support. The laser-induced signal is amplified using a CIVIDEC C2 current amplifier (2 GHz, 40 dB). After the amplification stage, the signal is digitized with an Agilent DSO 9254 Oscilloscope (2.5 GHz, 20 Gsa/s). More details about the setup can be found here \cite{CURRAS2022166530}. For these measurements, the intensity of the IR-laser was tuned to generate an equivalent charge of $\sim1\,MIPs$ with a laser beam spot size of around 20$\,\mu m$ in FWHM  to minimize the gain suppression effect \cite{CURRAS2022166530}. The measurements were performed at $-20\,^oC$ with the guard ring left floating. 

\begin{figure}[!t]
    \centering
  \begin{subfigure}[b]{0.49\textwidth}
         \centering
         \includegraphics[width=\textwidth]{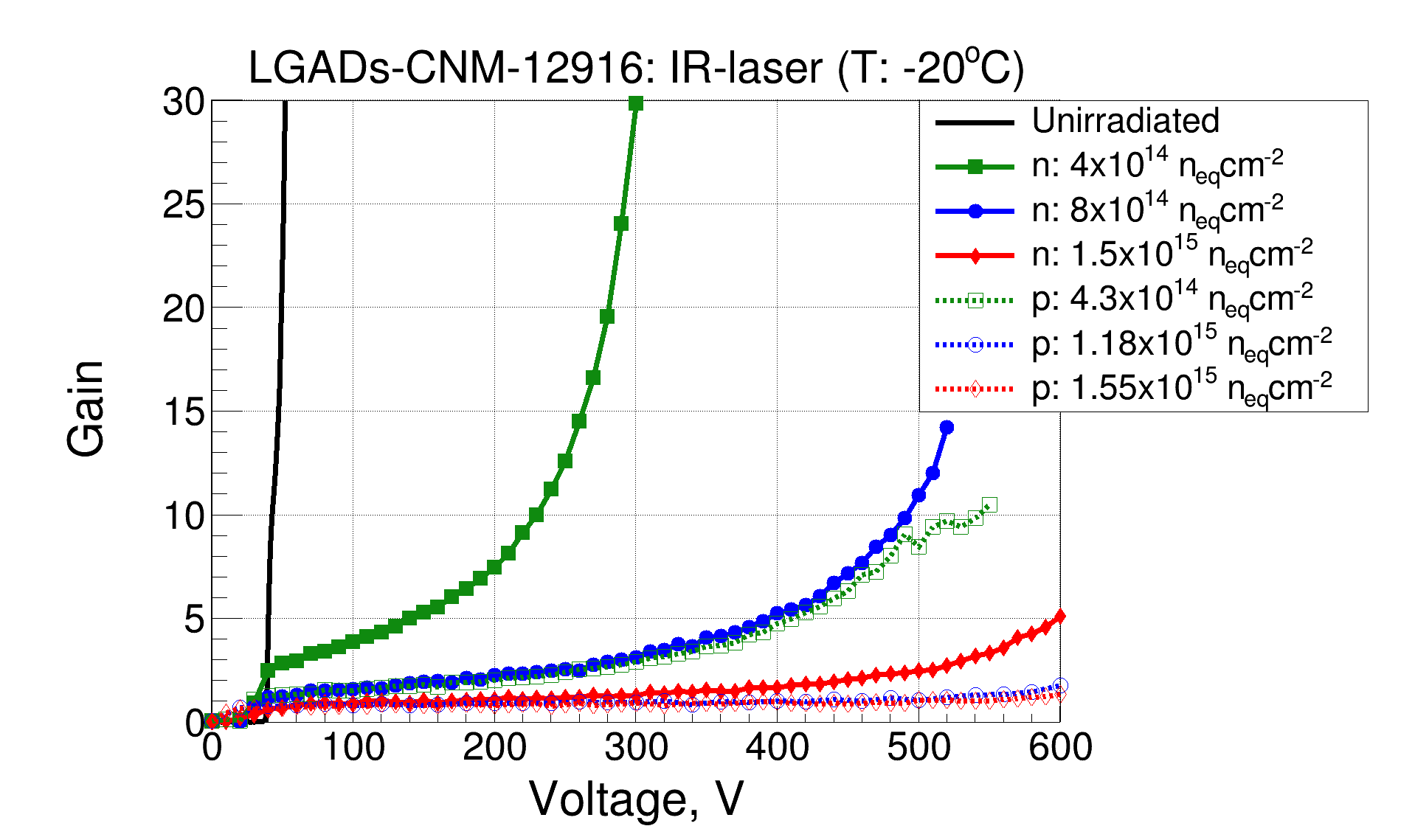}
         \caption{}
         \label{}
     \end{subfigure}
     \hfill
     \begin{subfigure}[b]{0.49\textwidth}
         \centering
         \includegraphics[width=\textwidth]{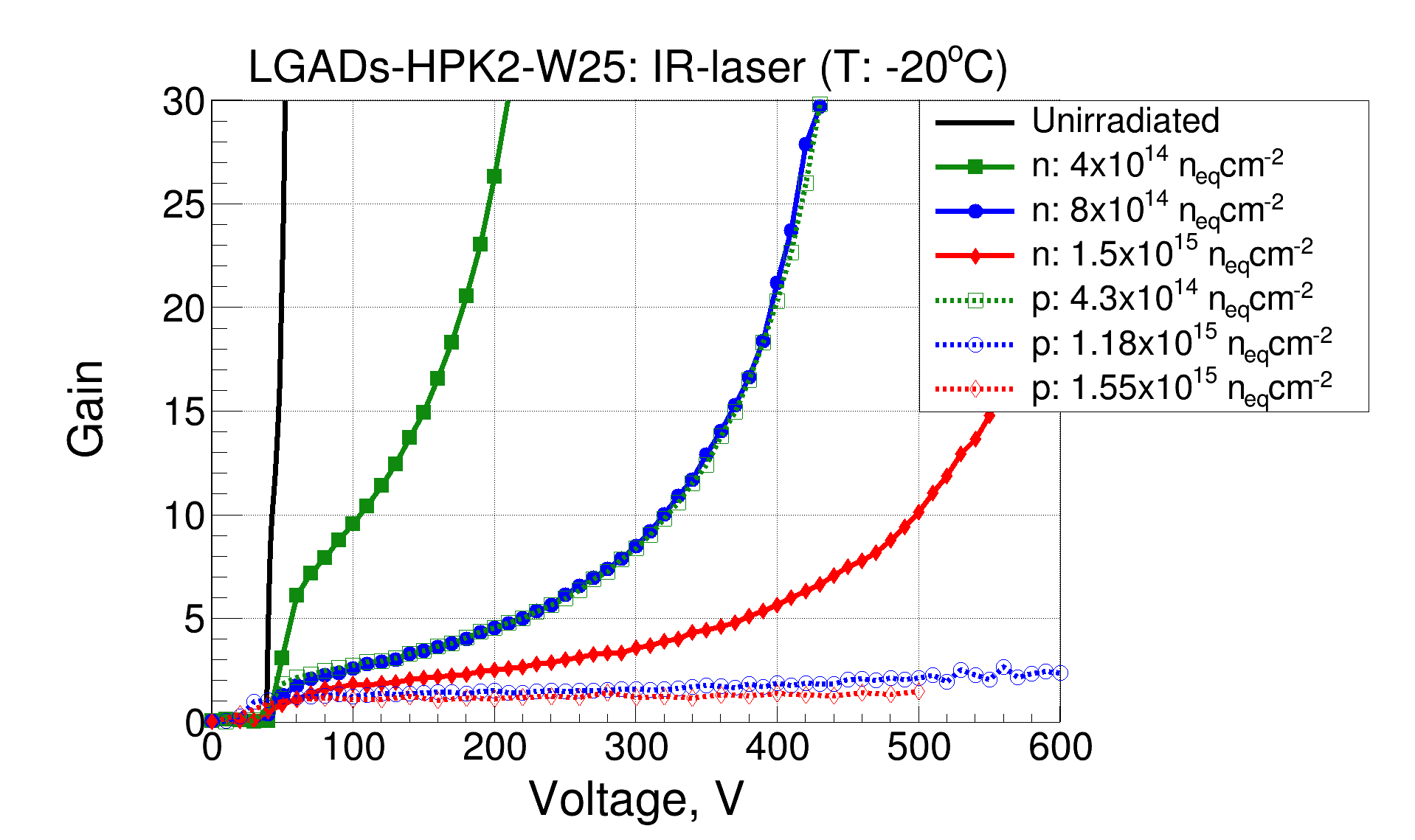}
         \caption{}
         \label{}
     \end{subfigure}
     \hfill
    \caption{Gain measurements using an IR-laser of the LGADs studied in this work. The charge measure in the LGAD was divided by the one measured in the equivalent unirradiated PIN to evaluate the gain in the CNM LGADs (a) and HPK ones (b).}
    \label{figure:5}
\end{figure}

The gain measurements are shown in \mbox{figure\,\ref{figure:5} }. The maximum bias voltage achieved was given by the IV measurements shown in the previous section, and in all the cases was kept below \mbox{600 V}. The gain was evaluated as the ratio between the LGAD collected charge ($CC_{LGAD}[V]$) and the equivalent unirradiated PIN collected charge after full depletion ($CC_{PIN}[V\ge V_{FD}]$), as a function of the reverse bias voltage (V) as expressed by:\\

\begin {equation}
\label{gain_formula}
  Gain[V] = \frac{CC_{LGAD}[V]}{CC_{PIN}[V\ge V_{FD}]}
\end{equation} \\

\section{Experimental results: gain layer degradation}

The results presented in this paper show a clear degradation of the LGADs properties with the irradiation fluence, being more severe in the case of the CNM LGADs than in the HPK ones, and both types of LGADs present more damage after the irradiation with \mbox{24\,GeV/c} protons than after the irradiation with neutrons when normalized to the \mbox{1 MeV} neutron equivalent fluence. These results can be seen in a qualitative way in the electrical properties. In \mbox{figure\,\ref{figure:3}} it is shown how the breakdown voltage increases with irradiation fluence, a clear indication of the degradation of the GL. Also, it is observed that the proton-irradiated LGADs show a higher breakdown voltage than the neutron-irradiated ones, implying more degradation in the GL. A similar conclusion can be extracted from the capacitance curves shown in \mbox{figure\,\ref{figure:4}}, where a clear decrease of the depletion voltage of the gain layer (V$_{GL}$) with the irradiation fluence is shown. Moreover, for the proton-irradiated LGADs, the decrease of (V$_{GL}$) is higher than for the neutron-irradiated ones. 
The gain measurements shown in \mbox{figure\,\ref{figure:5}} confirm these conclusions too. A higher degradation of the gain with the irradiation fluence in the CNM LGADs and also in both types of LGADs a higher degradation after the proton irradiation is observed.

To express these results in a quantitative way, we can evaluate the acceptor removal effect in the GL. The deactivation of the Boron (B) implanted in the GL with fluence, is linked with the reduction of the voltage needed to deplete the GL. If there is less active B inside the GL, the reverse bias voltage needed to fully deplete the GL is lower. Following this assumption, it is possible to extract the V$_{GL}$ from the data, and in consequence, the amount of active B remaining inside the GL. 

\begin{figure}[!t]
    \centering
  \begin{subfigure}[b]{0.31\textwidth}
         \centering
         \includegraphics[width=\textwidth]{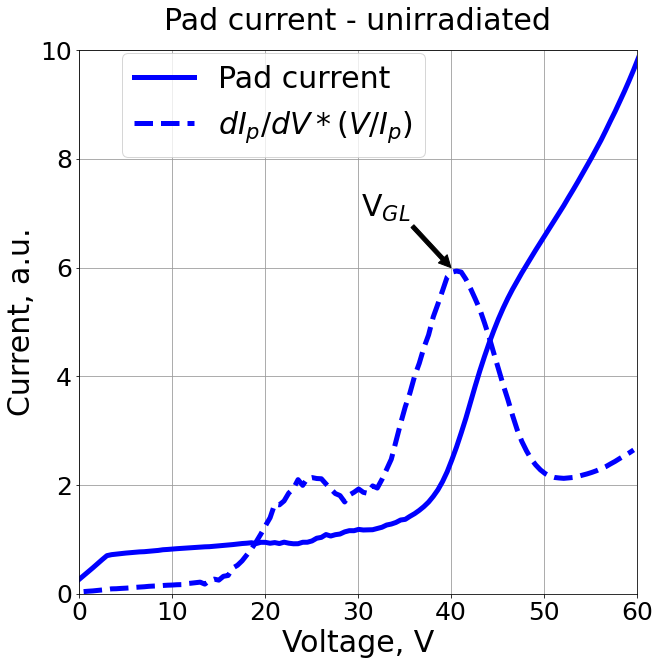}
         \caption{}
         \label{}
     \end{subfigure}
     \hfill
     \begin{subfigure}[b]{0.32\textwidth}
         \centering
         \includegraphics[width=\textwidth]{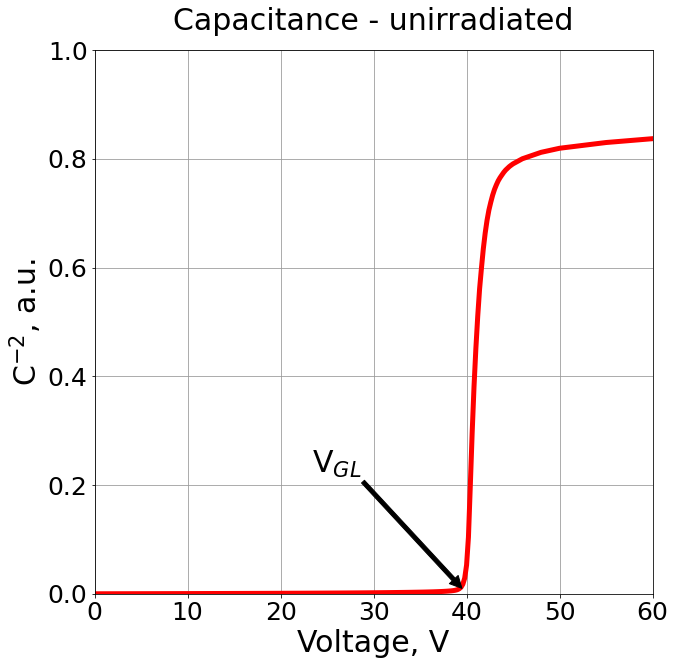}
         \caption{}
         \label{}
     \end{subfigure}
     \hfill
    \hfill
     \begin{subfigure}[b]{0.32\textwidth}
         \centering
         \includegraphics[width=\textwidth]{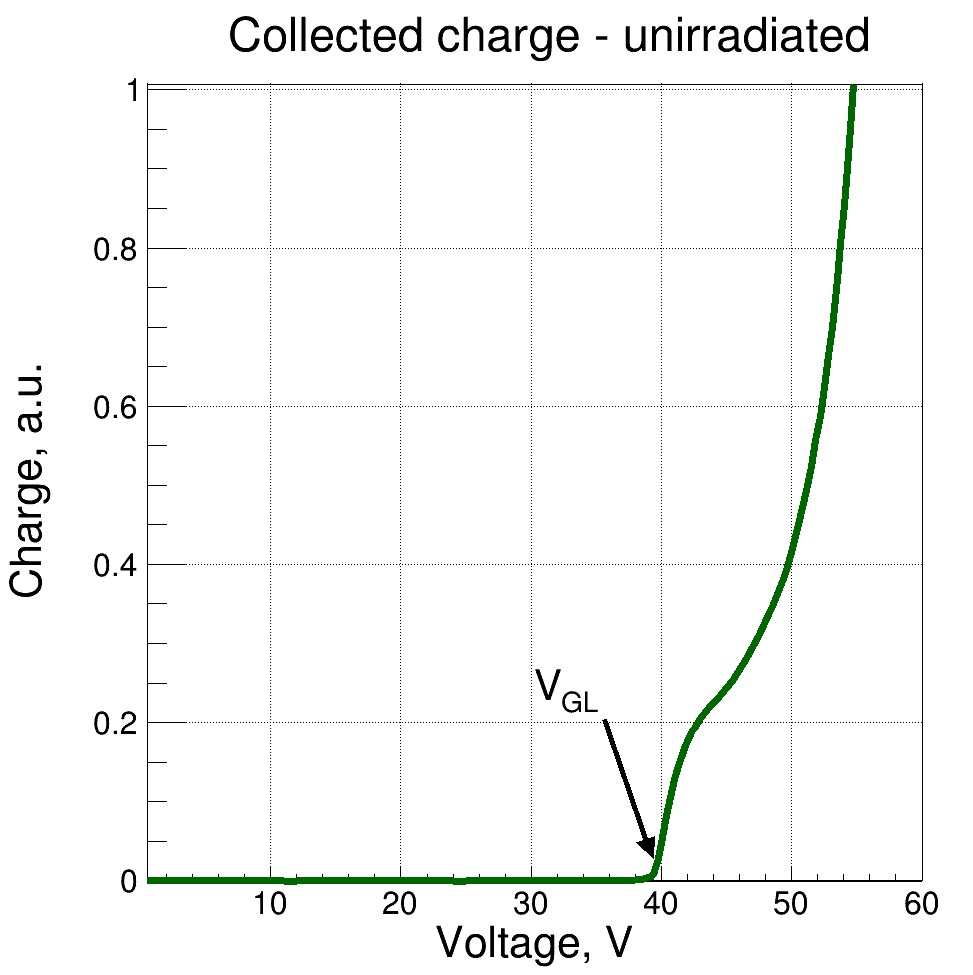}
         \caption{}
         \label{}
     \end{subfigure}
     \hfill
    \caption{Examples of the extraction of the $V_{GL}$ from the three different methods used in this work: (a) from the pud current curve, (b) from the capacitance curve, and, (c) from the collected charge curve.}
    \label{figure:6}
\end{figure}

The extraction of the V$_{GL}$ was done using three different methods that are depicted in \mbox{figure\,\ref{figure:6}}: figure (a) shows V$_{GL}$ in the pad current curves, figure (b) shows V$_{GL}$ in the inverse of the squared capacitance curves and figure (c) show the position of V$_{GL}$ in the collected charge measurement performed with the pulsed IR-laser. All these three methods give similar results within a \mbox{$\sim\,3\%$\, error} and an average value was taken to evaluate the acceptor removal effect. The initial acceptor removal is exponentially dependent on fluence, and it is defined in equation \ref{equation_C}, where c is the removal constant, N$_{x,0}$ initial doping concentration, and V$_{GL,0}$ the initial depletion voltage of the GL \cite{Kramberger_2015}.

\begin{figure}[!t]
    \centering
  \begin{subfigure}[b]{0.49\textwidth}
         \centering
         \includegraphics[width=\textwidth]{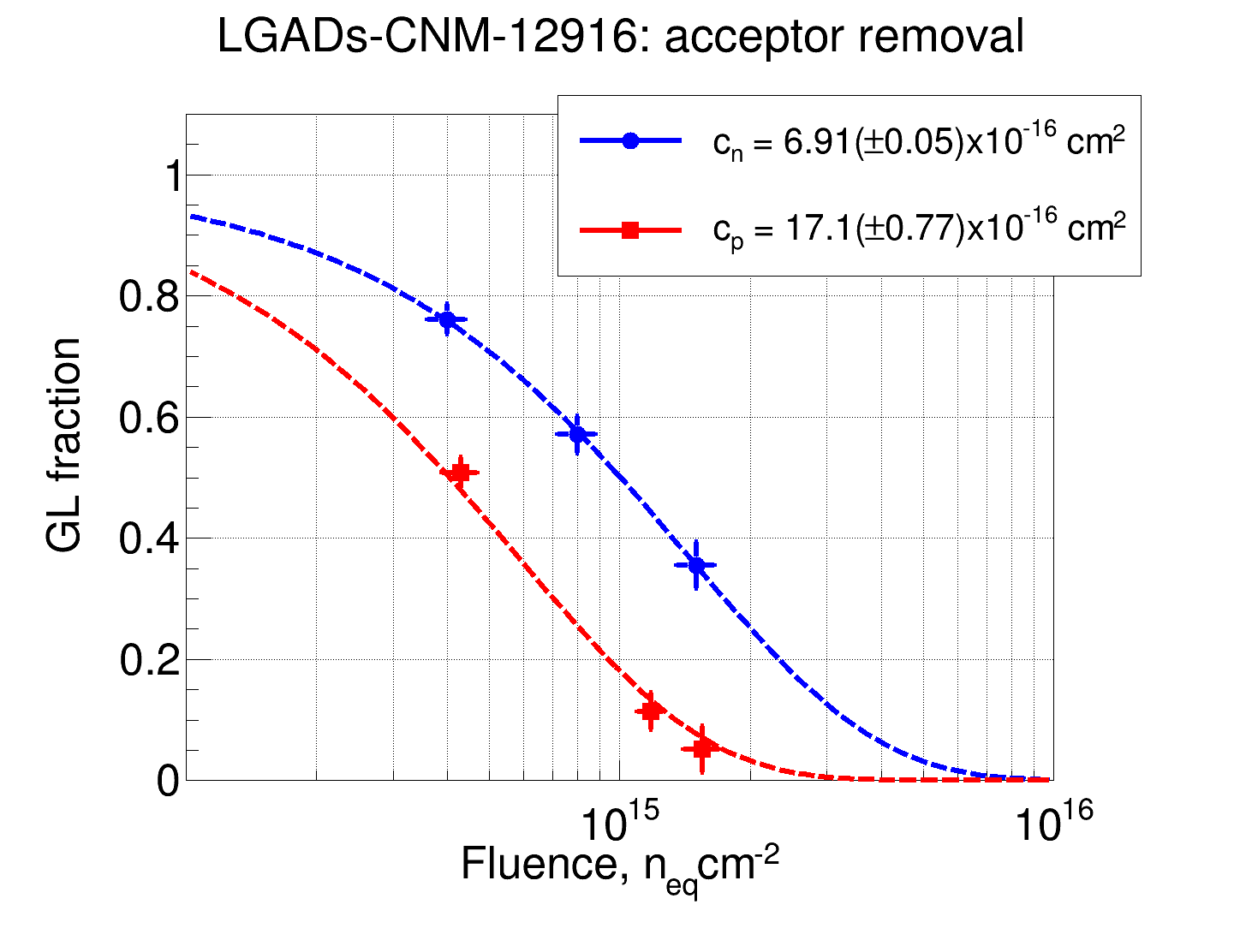}
         \caption{}
         \label{}
     \end{subfigure}
     \hfill
     \begin{subfigure}[b]{0.49\textwidth}
         \centering
         \includegraphics[width=\textwidth]{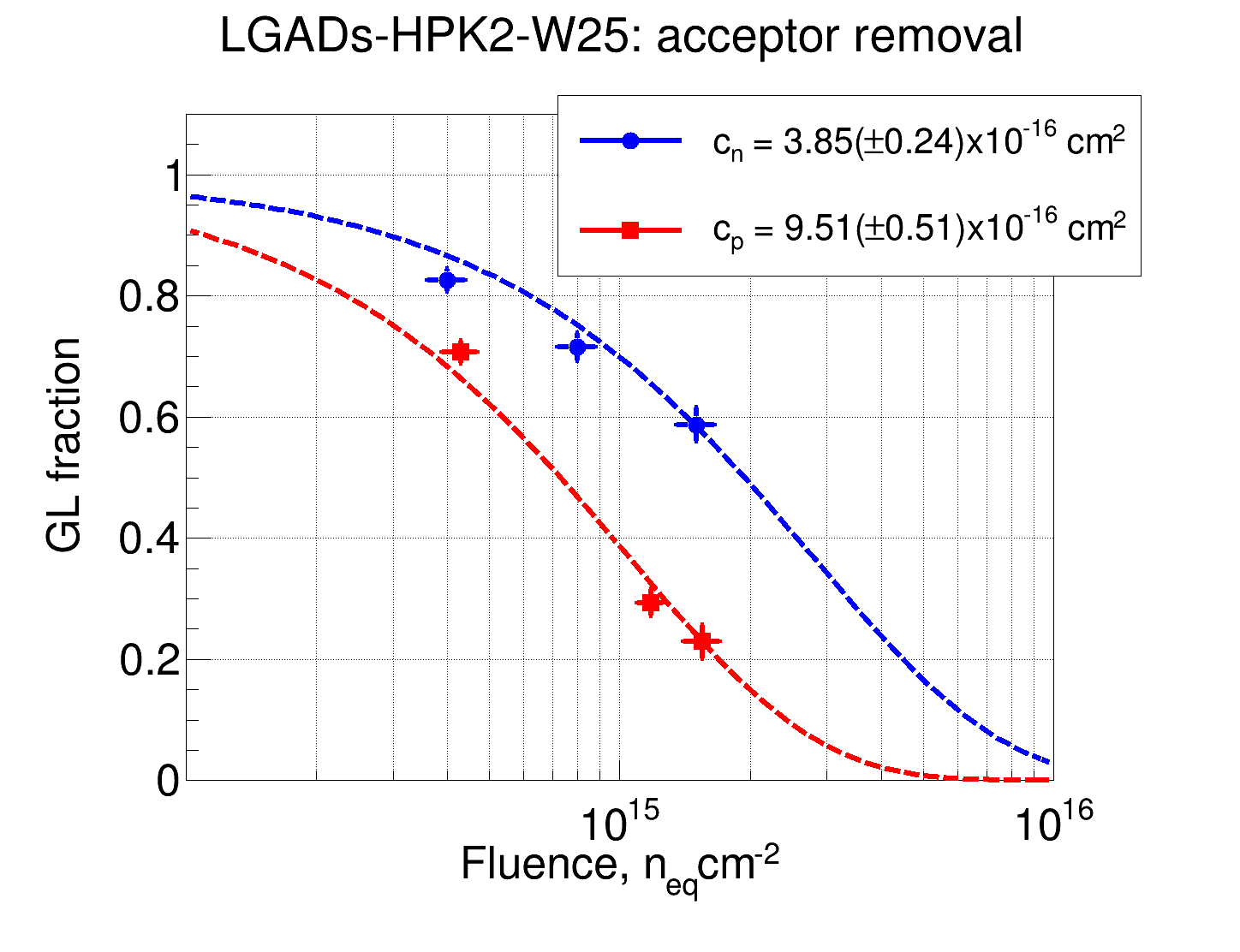}
         \caption{}
         \label{}
     \end{subfigure}
     \hfill
    \caption{Acceptor removal parameterization using equation \ref{equation_C} for the CNM LGADs (a) and for the HPK ones (b).}
    \label{figure:7}
\end{figure}

The GL fraction as a function of the irradiation fluence is shown in \mbox{figure\,\ref{figure:7}} for the CNM LGADs (a) and HPK LGADs (b). The GL fraction is calculated using the V$_{GL}$ values extracted from the previously mentioned methods as indicated in \mbox{equation \ref{equation_VGL}}. Also, in \mbox{figure\,\ref{figure:7}} the fitting curves of the acceptor removal (equation \ref{equation_C}) are shown with their respective values of the removal constant. For clarity, the removal constants values are shown in \mbox{table\,\ref{table_3}}. The results show in a consistent way that the removal constant for \mbox{24\,GeV/c} protons is $\sim\,2.5$ times higher than the one for neutrons when applying the normalization to \mbox{1 MeV} neutron equivalent fluence, and the removal constants in the CNM LGADs are $\sim\,1.8$ times higher than the ones in the HPK LGADs.

\small
\begin{equation}
\label{equation_C}
N_{x} = N_{x,0}\,\,exp(-c\, \Phi _{eq}) \,\, \Rightarrow \,\, V_{GL} \approx V_{GL,0}\,\,exp(-c\, \Phi _{eq})
\end{equation}
\normalsize

\small
\begin{equation}
\label{equation_VGL}
GL\,fraction = \frac{V_{GL}(\phi _{eq})}{V_{GL}(0)}
\end{equation}
\normalsize

\begin{table}[t]
\begin{center}
\caption{Table summarizing the acceptor removal constant (c) extracted from \mbox{figure\,\ref{figure:7}} using (equation \ref{equation_C}).\\} 
\label{table_3}
 \begin{tabular}{| l | c | c | c |} 
 \hline
  & $c_n [cm^2]$ & $c_p [cm^2]$ & $c_p/c_n$ \\ [0.7ex] 
 \hline\hline
 CNM LGADs & $6.91(\pm0.05)\times10^{-16}$ & $17.1(\pm0.77)\times10^{-16}$  & 2.475 \\[0.5ex] 
 \hline
  HPK LGADs & $3.85(\pm0.24)\times10^{-16}$ & $9.51(\pm0.51)\times10^{-16}$  & 2.470 \\[0.5ex] 
 \hline
 \hline
   $c_{CNM}/c_{HPK}$ & 1.795 & 1.798  &  -  \\[0.5ex] 
 \hline
\end{tabular}
\end{center}
\end{table}

It has been shown that the acceptor removal constant not only depends on the type of irradiation particle but also on the design of the GL. Different types of LGADs undergo less severe degradation depending on the shape, position, and concentration of the GL doping profile. Furthermore, the co-implantation of other elements, like carbon, influence the radiation resistance of the GL, see e.g. \cite{Curras:Vertex}. Therefore, comparing the acceptor removal constant obtained here with the values extracted from similar studies under the scope of the HL-LHC upgrade is interesting and it is needed to evaluate if the values extracted in this work are within the range of previously obtained data.

A study of the acceptor removal effect on IHEP-IMEv2 LGADs produced by the Institute of Microelectronics (Chinese Academy of Sciences) shows $c_{n}$ values in the range of $3.5-6.0\times10^{-16}cm^{2}$ after neutron irradiation \cite{9945985}. Different production runs of LGADs from HPK give $c_{n}$ values in the range of $3.15-5.2\times10^{-16}cm^{2}$ also after neutron irradiation and, $c_{p}$ values in the range of $6.5-7.0\times10^{-16}cm^{2}$ after irradiation with \mbox{70 MeV/c} protons \cite{JIN2020164611, A.Howard_RD50}. Different production runs of LGADs from CNM resulted in $c_{n}$ values in the range of $5.78-8.19\times10^{-16}cm^{2}$ after exposure to neutrons and $c_{p}$ values in the range of $18.7-19.6\times10^{-16}cm^{2}$ after exposure to \mbox{24 GeV/c} protons \cite{CURRAS2023, Vaguelis_RD50}. In all these cases, LGADs show an important improvement if carbon is co-implanted in the GL.

In conclusion, it can be stated that the value of the acceptor removal constant highly depends on the type of GL under study and that neutrons seem to be less damaging than protons when the fluence is normalized to the 1 MeV neutron equivalent fluence. Furthermore, although there are only a few radiation damage studies performed with protons, we observe that high-energy protons are much more damaging than low-energy protons. An observation that further measurements should consolidate.

\section{Summary and conclusions}
A gain layer degradation study of LGADs produced by Hamamatsu Photonics  and LGADs produced by Centro Nacional de Microelectrónica (CNM-IMB) was presented in this paper. The LGADs were exposed to reactor neutrons and \mbox{24\,GeV/c} protons at different fluences to evaluate the gain layer degradation under different types of particles. Electrical characterization and gain measurements with an IR laser were performed before and after irradiation in all the samples. An evaluation of the acceptor removal effect in the gain layer was conducted using different methods and two main conclusions can be extracted from this work: 
\begin{itemize}
  \item CNM LGADs produced with a shallow gain layer doping profile presented more degradation than the HPK LGADs produced with a deep gain layer doping profile. The acceptor removal constants (for protons and neutrons) gave a value in the CNM LGADs $\sim\,1.8$ times higher than in the HPK ones.
  \item The \mbox{24\,GeV/c} protons produced more damage in the gain layer than the neutrons in both types of LGADs. This was quantified with the acceptor removal constant, obtaining a ratio between protons and neutrons of $c_p/c_n\,\simeq\,2.5$ for both types of LGADs.
\end{itemize}

\acknowledgments
This work has been performed in the framework of the RD50 collaboration and the CERN EP R\&D programme on technologies for future experiments. This project has received funding from the European Union’s Horizon 2020 Research and Innovation programme under  GA no 101004761. We acknowledge the support from the Polish Ministry of Science and Higher Education and the National Science Centre, Poland (NCN), UMO-2018/31/B/ST2/03998.


\bibliographystyle{JHEP}
\bibliography{biblio.bib}

\end{document}